\documentstyle[pre,aps]{revtex}

\newcommand{\be}{\begin{equation}}
\newcommand{\ee}{\end{equation}}
\newcommand{\prt}{\partial}
\newcommand{\ep}{\varepsilon}

\begin{document}

\draft

\title{Transient Effect of Negative Electric Current in Irradiated 
Semiconductors}

\author{V.I. Yukalov$^{1,\dagger}$ and E.P. Yukalova$^2$}
\address{$^1$ Bogolubov Laboratory of Theoretical Physics \\
Joint Institute for Nuclear Research, Dubna 141980, Russia \\
and \\
Centre for Interdisciplinary Studies in Chemical Physics \\
University of Western Ontario, London, Ontario N6A 3K7, Canada \\
$^2$  Department of Computational Physics \\
Laboratory of Computing Techniques and Automation \\
Joint Institute for Nuclear Research, Dubna 141980, Russia \\
and \\
Department of Physics and Astronomy \\
University of Western Ontario, London, Ontario N6A 3K7, Canada\\
$\dagger$ e--mail: yukalov@thsun1.jinr.ru}

\maketitle

\begin{abstract}

The peculiarities of electric current are studied occurring in 
semiconductors with strongly nonuniform distribution of charge 
carriers. The formation of such nonuniformities and the regulation of 
carrier mobilities can be realized by means of external irradiation, for 
instance, charge--particle beams and laser irradiation. The transient 
effect of negative electric current is shown to arise under some specific 
conditions.

\end{abstract}

\section{Introduction}

The study of electric transport in semiconductors is important for 
describing and modelling different semiconductor devices [1,2]. The 
action of external irradiation can result in the formation in a 
semiconductor sample of nonuniform distributions of charge carriers. 
Thus, ion--beam irradiation leads to the formation of a dense layer of 
ions located at the distance of their mean free path from the surface.
Similar charged layers can be created by irradiating semiconductors with 
other beams of charged particles, say, electrons or positrons. If the 
injected charges can move, then the irradiated material behaves as an 
extrinsic semiconductor. The mobility of the charge carriers can be 
activated and regulated by involving additionally laser irradiation. Narrow 
laser beams can also be employed for creating nonuniform distributions of 
charge carriers.

Transport properties of semiconductors with essentially nonuniform 
distribution of carriers can be rather specific. For instance, in a 
sample, biased with an external constant voltage, the resulting electric 
current may turn against the latter [3,4]. Certainly, this can happen 
only as a short--time fluctuation after which the current turns back 
becoming positive [5,6]. The transient effect of negative electric 
current has been considered earlier [5,6] for simplified models. The aim 
of the present paper is to give a careful analysis of this effect under 
conditions typical of realistic semiconductor materials.

\section{Drift--Diffusion Equations}

Transport properties of semiconductors are usually described by the 
semiclassical drift--diffusion equations [1,2] consisting of the 
continuity equation
\be
\label{1}
\frac{\prt\rho_i}{\prt t} +\vec\nabla \cdot \vec j_i +
\frac{\rho_i}{\tau_i} = \xi_i \; ,
\ee
with the drift--diffusion current
\be
\label{2}
\vec j_i =\mu_i\rho_i\vec E - D_i \vec\nabla \rho_i\; ,
\ee
and of the Poisson equation
\be
\label{3}
\ep\vec\nabla \cdot \vec E = 4\pi (\rho_1 + \rho_2 ) \; .
\ee
Here $\rho_i=\rho_i(\vec r,t)$ is a charge density, 
$\vec E=\vec E(\vec r,t)$ is the electric field, and $\xi_i=\xi_i(\vec 
r,t)$, generation--recombination noise [7]. The considered semiconductor is
characterized by mobilities $\mu_i$, diffusion coefficients $D_i$, and a 
dielectric permittivity $\ep$. In what follows, two types of charge 
carriers are assumed, positive and negative, such that $\rho_1>0,\;
\mu_1>0$ and $\rho_2<0,\; \mu_2<0$. The total density of electric current 
writes
\be
\label{4}
\vec j_{tot} =\vec j_1 + \vec j_2 +\frac{\ep}{4\pi} 
\frac{\prt\vec E}{\prt t}\; ,
\ee
being the sum of the drift--diffusion current (2) and the displacement 
current.

Let us consider a plane device of area $A$ and width $L$, so that there 
is one space variable $x\in [0,L]$. Let the sample be biased with an 
external voltage $V_0$, which means that
\be
\label{5}
\int_0^L E(x,t)\; dx = V_0 \; .
\ee

Define the transit time
\be
\label{6}
\tau_0 \equiv \frac{L^2}{\mu V_0} \; , \qquad \mu\equiv \min\left\{ 
\mu_1 , \; |\mu_2|\right \} \; ,
\ee
and also introduce the following characteristic quantities
$$
\rho_0 \equiv \frac{Q_0}{AL} \; , \qquad Q_0 \equiv \ep AE_0\; , \qquad
E_0 \equiv \frac{V_0}{L} \; ,
$$
\be
\label{7}
j_0 \equiv \frac{Q_0}{A\tau_0} \; , \qquad D_0 \equiv \mu V_0\; , \qquad
\xi_0 \equiv \frac{\rho_0}{\tau_0} \; .
\ee
It is convenient to pass to dimensionless notation where the space 
variable $x$ is measured in units of $L$; time, in units of the transit 
time (6); and all other quantities in the corresponding units (7). Then 
the continuity equation (1) is written as
\be
\label{8}
\frac{\prt\rho_i}{\prt t} + \mu_i \frac{\prt}{\prt x} (\rho_i E) -
D_i\frac{\prt^2\rho_i}{\prt x^2} + \frac{\rho_i}{\tau_i} =\xi_i
\ee
and the Poisson equation (3) becomes
\be
\label{9}
\frac{\prt E}{\prt x}  = 4\pi (\rho_1 +\rho_2 ) \; ,
\ee
where the space and time variables are
\be
\label{10}
0 < x < 1 \; , \qquad t > 0 \; .
\ee
Equation (8) requires an initial and two boundary conditions. The first 
is given by the initial distribution of charge carriers
\be
\label{11}
\rho_i(x,0) = f_i(x) \qquad ( i=1,2) \; .
\ee
As the boundary conditions, we may accept the absence of diffusion 
through the semiconductor surface, which reads
\be
\label{12}
\frac{\prt}{\prt x}\rho_i(x,t) = 0 \qquad (x=0, \; x=1) \; .
\ee
The role of the boundary condition for Eq. (9) is played by the voltage 
integral (5) that in the dimensionless notation is
\be
\label{13}
\int_0^1 E(x,t)\; dx = 1 \; .
\ee

Note that by employing Eqs. (9) and (13), we may write the electric field 
in the form
\be
\label{14}
E(x,t) = 1 + 4\pi \left [ Q(x,t) - \int_0^1 Q(x,t)\; dx \right ] \; ,
\ee
where
$$
Q(x,t) = \int_0^x \left [ \rho_1(x',t) + \rho_2(x',t) \right ] \; dx' \; .
$$
As the total current (4), we have
\be
\label{15}
j_{tot} = \left ( \mu_1 E - D_1 \frac{\prt}{\prt x} \right )\rho_1 +
\left ( \mu_2 E - D_2 \frac{\prt}{\prt x} \right ) \rho_2 +
\frac{1}{4\pi}\;\frac{\prt E}{\prt t} \; .
\ee

The quantities whose time behaviour will be of interest for us are the 
electric current through the semiconductor sample
\be
\label{16}
J(t) \equiv \int_0^1 j_{tot}(x,t) \; dx
\ee
and the electric current at the left and right surfaces of the device,
\be
\label{17}
J(0,t) \equiv j_{tot}(0,t) \; , \qquad J(1,t) \equiv j_{tot}(1,t) \; .
\ee
The electric current (16), using Eq. (15), can be written as
$$
J(t) =\int_0^1\left [ \mu_1\rho_1(x,t) + \mu_2\rho_2(x,t) \right ] 
E(x,t)\; dx +
$$
\be
\label{18}
+ D_1\left [ \rho_1(0,t) - \rho_1(1,t)\right ]
+ D_2\left [ \rho_2(0,t) - \rho_2(1,t)\right ] \; ,
\ee
while for the currents (17) at the left and right surfaces, we get
\be
\label{19}
J(0,t) = J(t) + \int_0^1 [\gamma_1 Q_1(x,t) + \gamma_2 Q_2(x,t) ]\; dx\; ,
\ee
where
$$
Q_i(x,t) \equiv \int_0^x \rho_i(x',t)\; dx'\; , \qquad
\gamma_1 \equiv \frac{1}{\tau_i} \qquad (i=1,2) \; ,
$$
and, respectively,
\be
\label{20}
J(1,t) = J(0,t) - \int_0^1 [ \gamma_1\rho_1(x,t) + \gamma_2\rho_2(x,t) ]\;
dx \; .
\ee

\section{Negative Current}

In order to demonstrate that there exist conditions when the electric 
currents defined above can become negative, that is, directed against the 
applied voltage, let us take an illustrative example of narrow initial 
charge distributions which can be approximated by the form
\be
\label{21}
\rho_i(x,0) \equiv f_i(x) = Q_i\delta(x - a_i) \; .
\ee
Then from Eqs. (18)--(20), we have at the initial time
\be
\label{22}
J(0) = \mu_1 Q_1 E(a_1,0) + \mu_2 Q_2 E(a_2,0) \; ,
\ee
\be
\label{23}
J(0,0) = J(0) + \gamma_1 Q_1 (1 - a_1) + \gamma_2 Q_2 ( 1 - a_2) \; ,
\ee
\be
\label{24}
J(1,0) = J(0) - \gamma_1 Q_1a_1 - \gamma_2 Q_2 a_2 \; .
\ee
For the electric field (14), we get
$$
E(x,0) = 1 + 4\pi Q_1 [ a_1 - \Theta(a_1 - x ) ] + 4\pi Q_2 [a_2
- \Theta(a_2 - x ) ] \; ,
$$
where $\Theta(x)$ is the unit step function. Using this, for the total 
current (22), we find
$$
J(0) = \mu_1 Q_1 \left\{ 1 + 4\pi Q_1 \left ( a_1 - \frac{1}{2}\right ) +
4\pi Q_2 \left [ a_2 - \Theta(a_2 - a_1)\right ]\right \} +
$$
\be
\label{25}
+ \mu_2 Q_2 \left\{ 1 + 4\pi Q_2 \left ( a_2 - \frac{1}{2}\right ) +
4\pi Q_1 \left [ a_1 - \Theta(a_1 - a_2)\right ]\right \} \; .
\ee
The relation between the currents (23) and (24) has the form
$$
J(0,0) - J(1,0) = \gamma_1Q_1 + \gamma_2 Q_2 \; .
$$
In what follows, we shall consider two particular cases, when the initial 
charge distributions are located at the same place and when they are 
separated.

\subsection{Single--Layer Case}

Assume that both charge distributions (21) are located at the same place
\be
\label{26}
a_1 = a_2 \equiv a \; .
\ee
The initial electric field, then, is
\be
\label{27}
E(a,0) = 1 + 4\pi Q\left ( a - \frac{1}{2}\right ) \; , \qquad 
Q \equiv Q_1 + Q_2 \; .
\ee
The total current (25) becomes
\be
\label{28}
J(0) = ( \mu_1 Q_1 + \mu_2 Q_2 ) \left [ 1 + 4\pi Q\left ( a - \frac{1}{2}
\right )\right ] \; .
\ee
Since $\mu_iQ_i\geq 0$, we always have $\mu_1Q_1+\mu_2Q_2 >0$. Hence, the 
current (28) can be negative if
$$
a < \frac{1}{2} - \frac{1}{4\pi Q} \qquad (Q > 0 ) \; ,
$$
\be
\label{29}
a > \frac{1}{2} + \frac{1}{4\pi|Q|} \qquad (Q < 0) \; .
\ee
As far as $0<a<1$, inequalities (29) are possible for
\be
\label{30}
|Q| > \frac{1}{2\pi} \; .
\ee
Inverting Eq. (28), we may define the initial charge location
\be
\label{31}
a = \frac{1}{2} - \frac{1}{4\pi Q}\left [ 1 - 
\frac{J(0)}{\mu_1 Q_1 + \mu_2 Q_2}\right ]
\ee
as a function of the current $J(0)$. Measuring the latter gives us the 
location (31).

The current (23) at the left surface is negative under the condition
\be
a\left ( 4\pi Q - \frac{\gamma_1 Q_1 +\gamma_2 Q_2}{\mu_1Q_1 +\mu_2Q_2}
\right ) < 2\pi Q - 1 - \frac{\gamma_1Q_1 +\gamma_2Q_2}{\mu_1Q_1+\mu_2Q_2}
\; .
\ee
Also, measuring $J(0,0)$, we may define the location
\be
\label{33}
a = \frac{(2\pi Q -1)(\mu_1Q_1+\mu_2Q_2)-(\gamma_1Q_1+\gamma_2Q_2)+J(0,0)}
{4\pi Q(\mu_1Q_1 +\mu_2Q_2) - (\gamma_1Q_1 +\gamma_2Q_2)} \; .
\ee

The current (24) at the right surface becomes negative when
\be
\label{34}
a\left ( 4\pi Q - \frac{\gamma_1Q_1+\gamma_2Q_2}{\mu_1Q_1+\mu_2Q_2}\right )
< \;  2\pi Q - 1 \; .
\ee
The location of the initial charge layer can be defined through $J(1,0)$ as
\be
\label{35}
a = \frac{(2\pi Q-1)(\mu_1Q_1+\mu_2Q_2) + J(1,0)}{4\pi Q(\mu_1Q_1 +\mu_2Q_2)
-(\gamma_1Q_1+\gamma_2Q_2)}\; .
\ee

\subsection{Double--Layer Case}

Now consider the case when, at the initial time, two charge layers, 
described by the distributions (21), are separated in space so that
\be
\label{36}
a_1 = a < \; a_2 = 1 - a\; .
\ee
Then, substituting into the electric current (22) the electric fields
$$
E(a_1,0) = 1 - 2\pi Q_1 + 4\pi a (Q_1 - Q_2) \; ,
$$
$$
E(a_2,0) = 1 + 2\pi Q_2 + 4\pi a (Q_1 - Q_2 ) \; ,
$$
we have
\be
\label{37}
J(0) = \mu_1 Q_1 ( 1 -2\pi Q_1) + \mu_2Q_2 ( 1 +2\pi Q_2) +
4\pi a (Q_1 - Q_2) (\mu_1Q_1 + \mu_2 Q_2 )\; .
\ee
This current is negative if
\be
\label{38}
2a\left ( Q_1 - Q_2\right ) < \;
\frac{\mu_1Q_1^2 -\mu_2Q_2^2}{\mu_1Q_1+\mu_2Q_2} -\frac{1}{2\pi} \; .
\ee
In particular, when $Q_2=-Q_1$, Eq. (38) yields
$$
a < \frac{1}{4} - \frac{1}{8\pi Q_1} \; .
$$
This inequality, since $0<a<\frac{1}{2}$, gives $Q_1 >\frac{1}{2\pi}$.
The conditions for the currents (23) and (24) to be negative are
\be
\label{39}
\left [ 4\pi (Q_1 - Q_2) -
\frac{\gamma_1Q_1 -\gamma_2Q_2}{\mu_1Q_1 +\mu_2Q_2}\right ] a < \;
\frac{2\pi(\mu_1Q_1^2 -\mu_2Q_2^2)-\gamma_1Q_1}{\mu_1Q_1+\mu_2Q_2} -1
\ee
and, respectively,
\be
\label{40}
\left [ 4\pi (Q_1 - Q_2) -
\frac{\gamma_1Q_1 -\gamma_2Q_2}{\mu_1Q_1 +\mu_2Q_2}\right ] a < \;
\frac{2\pi(\mu_1Q_1^2 -\mu_2Q_2^2)+\gamma_2Q_2}{\mu_1Q_1+\mu_2Q_2} -1\; .
\ee
In this way, we see that each of the electric currents (22)--(24) can 
become negative at initial time, provided the corresponding conditions 
hold true.

\section{Qualitative Analysis}

To understand the general physical picture, we need to solve Eqs. (8) and
(9). The generation--recombination noise in Eq. (8) is usually modelled 
by the white Gaussian noise for which the stochastic averaging can be 
denoted by $\ll\ldots\gg$. This noise is defined by the mean
\be
\label{41}
\ll \xi_i(x,t)\gg = 0
\ee
and by the correlation function
\be
\label{42}
\ll \xi_i(x,t)\xi_j(x',t')\gg \; = \gamma_{ij}\delta(x-x')\delta(t-t') \; ,
\ee
where $\gamma_{ij}$ are the parameters characterizing the specific 
properties of the generation--recombination process.

An approximate solution of Eqs. (8) and (9) can be found by generalizing 
the method of scale separation [8-10] to the case of differential 
equations in partial derivatives. To this end, we have to find out which 
of the functions $E,\rho_1$ or $\rho_2$ could be considered as slow 
varying and, if so, with respect to what variables. In what follows we 
assume that
\be
\label{43}
\left | \int_0^\infty \ll \rho_1(x,t) +\rho_2(x,t) \gg dt\right |
< \infty \; ,
\ee
which will be confirmed a posteriori. Using Eqs. (9) and (43), we get
\be
\label{44}
\lim_{\tau\rightarrow\infty} \frac{1}{\tau}\int_0^\tau
\ll \frac{\prt}{\prt x} E(x,t) \gg dt = 0 \; ,
\ee
hence we can tell that the function $E$ is, on average, slowly varying in 
space. Then, because of the voltage integral (13), we have
\be
\label{45}
\int_0^1 \ll \frac{\prt}{\prt t} E(x,t) \gg dx = 0 \; ,
\ee
which suggests that $E$ is, on average, slowly varying in time. Thus, the 
function $E(x,t)$ can be treated as a quasi--invariant on average, with 
respect to both space and time. Keeping $E$ fixed in Eq. (8), we obtain a 
linear equation with respect to $\rho_i$, with constant coefficients. 
This equation, complimented by the initial conditions (11) and the 
boundary conditions (12), can be solved. The resulting solution looks a
little too cumbersome and we shall not write it down here in full, since 
our aim in this section is to give only a qualitative analysis for 
understanding the general physical picture. Therefore, we shall simplify 
the solution by considering a thick sample for which, instead of the 
boundary conditions (12), we may formally take
\be
\label{46}
\lim_{x\rightarrow \pm\infty} \ll \rho_i(x,t)\gg\; = 0 \; .
\ee
This simplification is equivalent to passing to an infinite sample, with 
the simultaneous continuation of $\rho_i$ outside the interval $[0,1]$ by 
setting $\rho_i=0$ for $x<0$ and $x>1$.

Then an approximate solution of Eq. (8) reads
\be
\label{47}
\rho_i = \rho_i^{reg} + \rho_I^{ran}\; ,
\ee
where the regular part
\be
\label{48}
\rho_i^{reg}(x,t) = \int_{-\infty}^{+\infty} G_i(x-x',t)\; f_i(x') \; dx'
\ee
is caused by the initial distribution $f_i(x)$, the Green function being
\be
\label{49}
G_i(x,t) = \frac{1}{2\sqrt{\pi D_i\; t}}\;  \exp\left\{ - 
\frac{(x-\mu_i E\; t)^2}{4 D_i\; t} - \gamma_i\; t\right\} \; ,
\ee
and the random part
\be
\label{50}
\rho_i^{ran}(x,t) = \int_0^t\int_{-\infty}^{+\infty} 
G(x-x',t-t')\xi_i(x',t')\; dx'\; dt'
\ee
is generated by the noise $\xi_i(x,t)$.  

The initial charge distribution can be modelled by the Gaussian
\be
f_i(x) = \frac{Q_i}{Z_i} \exp\left\{ - 
\frac{(x-a_i)^2}{2b_i}\right\} \; ,
\ee
where $0<a_i<1$ and
$$
Q_i =\int_0^1 f_i(x)\; dx \; , \qquad
Z_i =\int_0^1\exp\left\{ - \frac{(x-a_i)^2}{2b_i}\right\}\; dx \; .
$$
Strictly speaking, $f_i(x)$ is defined as zero for $x<0$ and $x>1$. But 
this restriction can be neglected in the thick--sample approximation 
which, by using the form (51) in the solution (48), yields
\be
\label{52}
\rho_i^{reg}(x,t) = \frac{Q_ib_i}{Z_i\sqrt{b_i^2 + 2D_i\;t}} \exp\left\{ - 
\frac{(x-\mu_i E\; t-a_i)^2}{2b_i^2 + 4D_i\; t} - \gamma_i\; t\right\} \; .
\ee

For the random solution (50), because of condition (41), we have
\be
\label{53}
\ll \rho_i^{ran}(x,t)\gg \; = 0 \; .
\ee
And from the definition (42), it follows
$$
\ll \rho_i^{ran}(x,t) \rho_j^{ran}(x',t)\gg\; = $$
\be
\label{54}
\int_0^t \frac{\gamma_{ij}}{2\sqrt{\pi(D_i+D_j)\; t}}\; \exp\left\{ -
\frac{[x-x'-(\mu_i-\mu_j)E\; t]^2}{4(D_i+D_j)\; t} - (\gamma_i +\gamma_j)\; t
\right \} \; dt \; ,
\ee
where $\gamma_{ij}$ are defined in Eq. (42). From here
\be
\label{55}
\lim_{t\rightarrow 0} \ll \rho_i^{ran}(x,t) \rho_j^{ran}(x',t) \gg\; = 0 \; .
\ee

Equations (52) and (53) show that $\ll\rho_i^{reg}\gg$ exponentially 
tends to zero as $t\rightarrow\infty$, thus, confirming inequality (43). 
The currents (18)--(20) are influenced by the noise through the 
correlators (54). The latter, according to Eq. (55), are small at short 
times. Therefore, at the beginning of the process, when $t\ll 1$, the 
role of noise is not important. This conclusion suggests that for 
considering transient effects, occurring at $t\ll 1$, the influence of the 
generation--recombination noise may be neglected.

\section{Numerical Solution}

To analyze more accurately the time behaviour of electric current, we 
have accomplished the numerical calculations of Eqs. (8) and (9) with
the initial conditions (11) and (51) and the boundary conditions (12) and
(13). In agreement with the previous section, the noise is neglected. All 
quantities are given in dimensionless units, as is explained in Sec. 2. 
Varying different parameters entering the problem, we fix $\mu_1=1$ and 
$Q_1=1$. We also keep in mind the relation $D_2=3D_1$ for the diffusion 
coefficients, typical of that for holes ($D_1$) and electrons ($D_2$). 
For short, we use the notation $\gamma_1=\gamma_2\equiv\gamma$ and 
$b_1=b_2\equiv b$. Figs. 1 and 2 present the results for the 
single--layer case, with $a_1=a_2\equiv a$; and Figs. 3 and 4, for the 
double--layer case, when $a_1\equiv a$, $a_2= 1-a$. The values of the
varying parameters are taken so that, when passing to dimensional units, 
they would correspond to the values characteristic for typical semiconductors 
[1,2]. The current $J(t)$, for $Q_1\geq |Q_2|$, lies always between 
$J(0,t)$ and $J(1,t)$, so that $J(1,t)\leq J(t)\leq J(0,t)$, as is clear 
from Eqs. (18)--(20). Therefore, we concentrate our attention on the 
behaviour of the limiting quantities $J(1,t)$ and $J(0,t)$. The general 
behaviour of the latter is in agreement with the qualitative analysis of 
Secs. 3 and 4. The principal thing which was impossible to notice in the 
qualitative analysis is that the electric current can become negative not 
at $t=0$, but at some finite time. Anyway, the occurrence of the negative 
current is a {\it transient effect} always happening at $t\ll 1$. The 
dependence of this effect on the physical parameters is thoroughly 
illustrated in the presented figures.

\vskip 5mm

{\bf Acknowledgement}

\vskip 2mm

We appreciate financial support from the University of Western Ontario, 
London, Canada, where this work was accomplished.

\vskip 2cm

\begin{center}
{\bf Figure Captions}
\end{center}

{\bf Fig. 1.} Single--layer case. 

\vskip 2mm

(a) The electric current $J(0,t)$ at the left surface of a semiconductor 
sample for the parameters $a=0.25,\; D_1=D_2=0,\; \gamma=1,\; \mu_2=-3$, for 
different negative charges: $Q_2=0$ (solid line), $Q_2=-0.25$ (long--dashed 
line), $Q_2=-0.5$ (short--dashed line), $Q_2=-0.75$ (dotted line), and 
$Q_2=-1$ (dashed--dotted line).

(a$'$) The electric current $J(1,t)$ at the right surface of semiconductor 
for the same parameters as in Fig. 1a.

(b) The electric current $J(0,t)$ at the left surface for the parameters
$a=0.25,\; \gamma=1,\; \mu_2=-3,\; Q_2=-0.5 $, and for different diffusion
coefficients: $D_1=0$ (solid line), $D_1=10^{-3}$ (long--dashed line),
$D_1=10^{-2}$ (short--dashed line), $D_1=10^{-1}$ (dotted line).

(b$'$) The electric current $J(1,t)$ at the right surface 
for the same parameters as in Fig. 1b.

(c) The electric current $J(0,t)$ at the left surface for the parameters
$a=0.25,\; \gamma=1,\; D_1=10^{-3},\; Q_2=-0.5$, and different mobilities
of the negative charge carriers:
$\mu_2=-10$ (solid line), $\mu_2=-5$ (long--dashed line),
$\mu_2=-3$ (short--dashed line).

(c$'$) The electric current $J(1,t)$ at the right surface 
for the same parameters as in Fig. 1c.

\vskip 1cm

{\bf Fig. 2.} Single--layer case. 

\vskip 2mm

The electric currents at the left, $J(0,t)$ (solid line), and at the right, 
$J(1,t)$ (long--dashed line), surfaces for the diffusion coefficient 
$D_1=10^{-3}$ and different sets of other parameters: 

(a) $a=0.25,\; \gamma=1,\; \mu_2=-3,\; Q_2=-0.5$.

(b) $a=0.35,\; \gamma=1,\; \mu_2=-3,\; Q_2=-0.5$.

(c) $a=0.05,\; \gamma=10,\; \mu_2=-3,\; Q_2=-0.5$.

(d) $a=0.25,\; \gamma=10,\; \mu_2=-3,\; Q_2=-0.5$.

(e) $a=0.35,\; \gamma=10,\; \mu_2=-3,\; Q_2=-0.5$.

(f) $a=0.25,\; \gamma=0.1,\; \mu_2=-3,\; Q_2=-0.5$.

(g) $a=0.25,\; \gamma=10,\; \mu_2=-10,\; Q_2=-0.5$.

(h) $a=0.35,\; \gamma=10,\; \mu_2=-3,\; Q_2=-1$.

\vskip 1cm

{\bf Fig. 3.} Double--layer case. 

\vskip 2mm

(a) The electric current $J(0,t)$ at the left surface of a semiconductor 
sample for the parameters $a=0.1,\; \gamma=1,\; \mu_2=-3,\; Q_2=-1$, and 
different diffusion coefficients: $D_1=0$ (solid line), $D_1=10^{-3}$ 
(long--dashed line), $D_1=10^{-2}$ (short--dashed line), $D_1=10^{-1}$ 
(dotted line).

(a$'$) The electric current $J(1,t)$ at the right surface 
for the same parameters as in Fig. 3a.

(b) The left--surface current $J(0,t)$ for the parameters 
$a=0.1,\; \gamma=1,\; \mu_2=-3,\; D_1=10^{-3}$, and different initial
charges: $Q_2=0$ (solid line), $Q_2=-0.25$ (long--dashed line),
$Q_2=-0.5$ (short--dashed line), $Q_2=-0.75$ (dotted line), and $Q_2=-1$
(dashed--dotted line).

(b$'$) The right--surface current $J(1,t)$ for the same parameters as in 
Fig. 3b.

(c) The left--surface current $J(0,t)$ for the parameters
$a=0.1,\; \gamma=1,\; D_1=10^{-3},\; Q_2=-1$, and different mobilities:
$\mu_2=-10$ (solid line), $\mu_2=-5$ (long--dashed line),
$\mu_2=-3$ (short--dashed line).

(c$'$) The right--surface current $J(1,t)$ for the same parameters as in 
Fig. 3c. For these parameters the left--and right--surface currents are 
indistinguishable, so the longer time interval is presented here.

(d) The left--surface current $J(0,t)$ for the parameters 
$a=0.25,\; \mu_2=-10,\; D_1=10^{-3},\; Q_2=-0.1$, and different relaxation 
widths:  $\gamma=25$ (solid line), $\gamma=10$ (long--dashed line),
$\gamma=1$ (short--dashed line).

(d$'$) The right--surface current $J(1,t)$ for the same parameters as in 
Fig. 3d.

\vskip 1cm

{\bf Fig. 4.} Double--layer case. 

\vskip 2mm

The electric currents at the left, $J(0,t)$ (solid line), and the right, 
$J(1,t)$ (long--dashed line), surfaces for the diffusion coefficient 
$D_1=10^{-3}$ and different sets of other parameters: 

(a) $a=0.1,\; \gamma=1,\; \mu_2=-3,\; Q_2=-0.5$.

(b) $a=0.1,\; \gamma=10,\; \mu_2=-3,\; Q_2=-0.5$.

(c) $a=0.1,\; \gamma=10,\; \mu_2=-3,\; Q_2=-1$.

(d) $a=0.25,\; \gamma=10,\; \mu_2=-3,\; Q_2=-0.25$.

(e) $a=0.25,\; \gamma=10,\; \mu_2=-10,\; Q_2=-0.25$.

(f) $a=0.25,\; \gamma=10,\; \mu_2=-10,\; Q_2=-0.1$.

(g) $a=0.25,\; \gamma=100,\; \mu_2=-10,\; Q_2=-0.1$.


\begin{references}

\bibitem{1} Snowden, C.M., 1986, {\it Introduction to Semiconductor 
Device Modelling} (Singapore: World Scientific).

\bibitem{2} Seeger, K., 1989, {\it Semiconductor Physics} (Berlin: Springer).

\bibitem{3} Rudenko, A.I. and Yukalov, V.I., 1981, in {\it Investigation 
of Surface and Volume Properties of Solids by Particle Interactions}, 
Ryazanov, M.I., Ed. (Moscow: Energoizdat), p.78.

\bibitem{4} Yukalov, V.I., 1985, {\it JINR Rapid Commun.}, {\bf 7}, 51.

\bibitem{5} Yukalov, V.I. and Yukalova, E.P., 1997, {\it Phys. Lett. A}, 
{\bf 236}, 113.

\bibitem{6} Yukalov, V.I. and Yukalova, E.P., 1997, {\it Laser Phys.}, 
{\bf 7}, 1076.

\bibitem{7} Van der Ziel, A., 1986, {\it Noise in Solid State 
Devices and Circuits} (New York: Wiley).

\bibitem{8} Yukalov, V.I., 1993, {\it Laser Phys.}, {\bf 3}, 870.

\bibitem{9} Yukalov, V.I., 1995, {\it Laser Phys.}, {\bf 5}, 970.

\bibitem{10} Yukalov, V.I., 1996, {\it Phys. Rev. B}, {\bf 53}, 9232.

\end{references}
\end{document}